\documentclass{ws-procs975x65}     %%uses ...my.sty file copied to 65.sty %%ws prc 2015 sty 

\usepackage{
slashed,yhmath,url}

\begin{document} 

%\addtolength{\oddsidemargin}{-0.5in} %%for arxiv only!!
%	\addtolength{\evensidemargin}{-0.5in} %%for arxiv only!!
 %    \addtolength{\topmargin}{-0.5in}%%for arxiv only!!
  %    \addtolength{\textheight}{-1in}%%for arxiv only!!
  %     \addtolength{\textwidth}{1.75in}
%================================================================
%================================================================

%\wstoc{On the effects of quantum spin connection foam in the Solar system, galaxies, and the Universe}{I.V. Kanatchikov and V.A. Kholodnyi}

\title{
 %On the effects of quantum spin connection foam in the Solar system, galaxies, and the Universe }
%\footnotesize 
 %\small EFFECTS OF QUANTUM SPIN CONNECTION FOAM IN \\\vspace{4pt} THE SOLAR SYSTEM, GALAXIES, AND THE UNIVERSE\footnote{
 %change 2026 
 Effects of Quantum Spin-Connection Foam in the Solar System, Galaxies, and the Universe\footnote{
%Based on the talk at MG17, Pescara, Italy, 2024 
To appear in: {\em The Seventeenth Marcel Grossmann Meeting
On Recent Developments in Theoretical and Experimental General Relativity, Astrophysics, and Relativistic Field Theories, 
 %Proceedings of the Seventeenth Marcel Grossmann Meeting  (on Recent Developments in Theoretical and Experimental General  Relativity, Gravitation, and Relativistic Field Theories), 
  Pescara 7-12 July 2024}, edited by G. Vereshchagin and R. Ruffini, \url{https://doi.org/10.1142/14814} | October 2026.}}

\author{Igor V. Kanatchikov$^{*,**}$ and Valery A. Kholodnyi$^{\star, \star\star}$}
 %\aindx{Kanatchikov, I. and Kholodnyi, V.}
\address{$^*$National Quantum Information Center in Gdansk (KCIK), Gda\'nsk, 80-309, Poland \\
$^{**}$IAS-Archimedes Project, Saint-Rapha\"{e}l 83700, France \\
$^\star$Wolfgang-Pauli-Institute, Oskar-Morgenstern-Platz 1, 1090 Vienna, Austria \\
$^{\star \star}$Unyxon, Woodforest TX, USA 
 %\email{kanattsi@gmail.com}}
 }

\begin{abstract} 
We argue that effects of the quantum spin-connection foam, which describes quantum gravity according to the precanonical quantization of General Relativity, may already be observed in the form of the small cosmological constant and a modification of Newtonian dynamics at small accelerations, manifested in the flat rotation curves of galaxies. We obtain a modification of the Newtonian potential that takes into account the existence of a fundamental small acceleration scale, $a_* = 8\pi G\hbar\varkappa$, where $\varkappa$ is a parameter with the dimensions of inverse spatial volume that appears on dimensional grounds. The connection between $\varkappa$ and the hadronic scale of the mass gap in the pure Yang–Mills sector of the Standard Model leads to an estimated value of $a_*$ compatible with the Milgromian acceleration scale in MOND. The connection between $a_*^2$ and the cosmological constant leads to a realistic value of the latter. Milgromian MOND, together with a theoretically distinct interpolating function, is derived under the assumption that classical dynamics is modified by the mean-field acceleration calculated from the simplest solution of precanonical quantum gravity in the nonrelativistic approximation. We also indicate that the effects of Newtonian dynamics modified by the spin-connection foam may be observable in the Solar System and even in laboratory experiments. 
%MG17 published 
%We assert that the effects of the quantum spin connection foam, which describes quantum gravity according to the precanonical quantization of General Relativity, are already being observed in the form of the small cosmological constant and the modification of the Newtonian dynamics at small accelerations, which manifests itself in flat rotation curves of galaxies. We obtain the modification of the Newtonian potential that takes into account the existence of the fundamental small acceleration scale $a_*=8\pi G \hbar \varkappa$, where $\varkappa$ is the parameter of the dimension of the inverse spatial volume, which appears on dimensional grounds. The connection between $\varkappa$ and the hadronic mass gap in the pure Yang-Mills sector of the Standard Model leads to the estimated value of $a_*$ compatible with the Milgromian acceleration in MOND. The connection between $a^2_*$ and the cosmological constant leads to the realistic value of the latter. The Milgromian MOND, together with the theoretically distinct interpolating function, are derived under the assumption that the classical dynamics is altered by the mean field acceleration calculated from the most straightforward solution of precanonical quantum gravity in the nonrelativistic approximation. We also indicate that the effects of the Newtonian dynamics modified by the spin connection foam may be observed in the Solar system and even in a laboratory.
\end{abstract}
\bodymatter

\newcommand{\BibTeX}{B\kern-0.1emi\kern-0.017emb\kern-0.15em\TeX}
\newcommand{\XYpic}{$\mathrm{X\kern-0.3em\raisebox{-0.18em}{Y}}$-$\mathrm{pic}\,$}
\newcommand{\TexnicCenter}{\TeX nicCenter}

%%%Clifford algebra macros
\newcommand{\cl}{C \kern -0.1em \ell}  %%Clifford algebra

\newcommand{\w}{\wedge}
\newcommand{\bigw}{\bigwedge}
\newcommand{\DDW}{\mathbin{\dot\wedge}}
\newcommand{\DDWedge}{\mathbin{\dot\wedge}}

\newcommand{\by}{\mathbf{y}}
\newcommand{\ba}{\mathbf{a}}
\newcommand{\bb}{\mathbf{b}}

\newcommand{\BF}{\mathbb{F}}
\newcommand{\BZ}{\mathbb{Z}}
\newcommand{\BR}{\mathbb{R}}
\newcommand{\BC}{\mathbb{C}}
%%%\newcommand{\BH}{\mathbb{H}}

%\newcommand{\ed}{\end{document}}

%%%% NEW   
 %\usepackage{yhmath} %%the simplest solution for wider \widehat{}!!!!!
  
   %\usepackage[lite]{mtpro2}  %%really wide hat \widehat{},    instructions provided at pctex.com/mtpro2.html

%%%%%%%%%%%%%%%%%%%%%%%%%%%%%%%%%%%%%%%%%%%%
%% MY NEWCOMMANDS  %%%%%%%%%%%%%%%%%%%%%%%%
%%%%%%%%%%%%%%%%%%%%%%%%%%%%%%%%%%%%%%%%%%%%

\newcommand{\beq}{\begin{equation}}
\newcommand{\eeq}{\end{equation}}
\newcommand{\beqa}{\begin{eqnarray}}
\newcommand{\eeqa}{\end{eqnarray}}
\newcommand{\nn}{\nonumber}
\newcommand{\half}{\frac{1}{2}}
\newcommand{\xt}{\tilde{X}}
%%%%rm for Minsk%%\newcommand{\uind}[2]{^{#1_1 \, ... \, #1_{#2}} }
%%%%  \newcommand{\lind}[2]{_{#1_1 \, ... \, #1_{#2}} }
\newcommand{\com}[2]{[#1,#2]_{-}} 
\newcommand{\acom}[2]{[#1,#2]_{+}} 
\newcommand{\compm}[2]{[#1,#2]_{\pm}}

\newcommand{\lie}[1]{\pounds_{#1}}
\newcommand{\co}{\circ}
\newcommand{\sgn}[1]{(-1)^{#1}}
\newcommand{\lbr}[2]{ [ \hspace*{-1.5pt} [ #1 , #2 ] \hspace*{-1.5pt} ] }
\newcommand{\lbrpm}[2]{ [ \hspace*{-1.5pt} [ #1 , #2 ] \hspace*{-1.5pt}
 ]_{\pm} }
\newcommand{\lbrp}[2]{ [ \hspace*{-1.5pt} [ #1 , #2 ] \hspace*{-1.5pt} ]_+ }
\newcommand{\lbrm}[2]{ [ \hspace*{-1.5pt} [ #1 , #2 ] \hspace*{-1.5pt} ]_- }

%%%%%%%%% adapted to 11pt %%%%%%%%% 
%\newcommand{\pbr}[2]{ \{ \hspace*{-3.3pt} [ #1 , #2\hspace*{1.4 pt} ] 
%\hspace*{-3.3pt} \} }
%%\newcommand{\nbr}[2]{ [ \hspace*{-2.0pt} [ #1 , #2\hspace*{1.4 pt} ] 
%%\hspace*{-2.0pt} ] }
%\newcommand{\nbr}[2]{ [ \hspace*{-1.5pt} [ #1 , #2 \hspace*{0.pt} ] 
%\hspace*{-1.3pt} ] }

%%%% adapted to AIP conf sty %%%%
\newcommand{\pbr}[2]{ \{ \hspace*{-2.6pt} [ #1 , #2\hspace*{1.4 pt} ] 
\hspace*{-2.6pt} \} }
\newcommand{\nbr}[2]{ [ \hspace*{-1.5pt} [ #1 , #2 \hspace*{0.pt} ] 
\hspace*{-1.3pt} ] }

\newcommand{\we}{\wedge}
\newcommand{\nbrpq}[2]{\nbr{\xxi{#1}{1}}{\xxi{#2}{2}}}
\newcommand{\lieni}[2]{$\pounds$${}_{\stackrel{#1}{X}_{#2}}$  }

\newcommand{\rbox}[2]{\raisebox{#1}{#2}}
\newcommand{\xx}[1]{\raisebox{1pt}{$\stackrel{#1}{X}$}}
\newcommand{\xxi}[2]{\raisebox{1pt}{$\stackrel{#1}{X}$$_{#2}$}}
\newcommand{\ff}[1]{\raisebox{1pt}{$\stackrel{#1}{F}$}}
\newcommand{\dd}[1]{\raisebox{1pt}{$\stackrel{#1}{D}$}}
\newcommand{\der}{\partial}

\newcommand{\Lm}{\bigwedge^*}

\newcommand{\inn}{\hspace*{2pt}\raisebox{-1pt}{\rule{6pt}{.3pt}\hspace*
{0pt}\rule{.3pt}{8pt}\hspace*{3pt}}}
\newcommand{\sro}{Schr\"{o}dinger\ }
\newcommand{\vol}{\omega}%%{\widetilde{vol}}                              
               \newcommand{\dvol}[1]{\der_{#1}\inn \vol}

\newcommand{\bd}{\mbox{\bf d}}
\newcommand{\bder}{\mbox{\bm $\der$}}
\newcommand{\bI}{\mbox{\bm $I$}}

\newcommand{\be}{\beta} 
\newcommand{\gab}{\bar{\gamma}} 
\newcommand{\ga}{\gamma} 
\newcommand{\de}{\delta} 
\newcommand{\gmu}{\bar{\gamma}^\mu}
\newcommand{\gnu}{\bar{\gamma}^\nu}
\newcommand{\ka}{\varkappa}
\newcommand{\hka}{\hbar\varkappa}

\newcommand{\lapl}{\bigtriangleup}
\newcommand{\psib}{\overline{\psi}}
\newcommand{\Psib}{\overline{\Psi}}
\newcommand{\Phib}{\overline{\Phi}}
\newcommand{\derts}{\stackrel{\leftrightarrow}{\der}}
\newcommand{\deltab}{\overline{\delta}}

\newcommand{\pib}{\overline{\pi}}
\newcommand{\Cb}{\overline{C}}
\newcommand{\ub}{\overline{u}}

\newcommand{\bx}{{\mathbf{x}}}
\newcommand{\bk}{{\bf k}}
\newcommand{\bq}{{\bf q}}
\newcommand{\bee}{{\bf e}}

\newcommand{\omk}{\omega_{\bf k}} 
\newcommand{\lpl}{\ell}
\newcommand{\zb}{\overline{z}} 

\newcommand{\dv}{\mbox{\sf d}}

\newcommand{\BPsi}{{\bf \Psi}} %Stopped  to work with AIP proc. style!
       %\newcommand{\BPsi}{\boldsymbol\Psi} 
       %\newcommand{\BPsi}{\mathbf{\Psi}} 
   
 %Imitation of BOLD capital Psi:   %for AIP proc sty!!!    
 %\newcommand{\BPsi}{\Psi\!\!\!\!\!\!\!\Psi} %Almost works 

\newcommand{\BPhi}{{\bf \Phi}} 
\newcommand{\BXi}{{\bf \Xi}} 
\newcommand{\BH}{{\bf H}} 
\newcommand{\BS}{{\bf S}} 
\newcommand{\BN}{{\bf N}}

\newcommand{\hy}{\hat{y}}
\newcommand{\hP}{\hat{P}}
\newcommand{\hH}{\what{H}}

\newcommand{\myvdots}{\mbox{\raisebox{-2pt}{$\vdots$}}}

\newcommand{\fbar}{\bar{f}}

%\DeclareMathOperator{\Tr}{Tr} %{\mathfs{Tr}}
%\DeclareMathOperator{\1F1}{{}_1F_1} %{\mathfs{Tr}}

%%%%%%%%%%%%%%%%%%%%%%%%%%%%%%%%%%%%%%%%%%%%%%%%%%%%%%
%%%% MLADENOV"S DEFS %%%%%
\newcommand{\rd}{\mathrm{d}} % Roman d for differential
\newcommand{\re}{\mathrm{e}} % Roman e for exponential
\newcommand{\ri}{\mathrm{i}} % Roman i for imaginary number
\def\semicolon{\nobreak\mskip2mu\mathpunct{}\nonscript\mkern-\thinmuskip{;}
\mskip6muplus1mu\relax} % This defines the semicolon command 
%%%%%%%%%%%%%%%%%%%%%%%%%%

\newcommand{\omegab}{\bar{\omega}}
\newcommand{\gammab}{\bar{\gamma}}
\newcommand{\betab}{\bar{\beta}} 
\newcommand{\ugamma}{\underline{\gamma}}

\newcommand{\abar}{\bar{a}}

\newcommand{\ddd}{\mathbf{d}}

%%2021
\newcommand{\fe}{\mathfrak{e}}
\newcommand{\fp}{\mathfrak{p}}
\newcommand{\fC}{\mathfrak{C}}

%%%%%%%%%%%%%%%%%%%%%%%%%%%%%%%%%%%%%%%%%%%%%%%%%%%%%%% 
%%%%%%%%%%%%%%%%%%%%%%%%%%%%%%%%%%%%%%%%%%%%%%%%%%%%%%%  

%\begin{document}

\section{Introduction}

Recently, there has been an increasing interest in discussing different effects of quantum gravity that 
could be potentially observed beyond the realm of the Planck scale \cite{ameli1,menezes,addazi,alonso,freidel}.  
Here, we argue that the effects of quantum gravity are possibly 
 already %%actually %being  
 observed in the form of the cosmological constant, the Milgromian acceleration and the certain 
 modification of the Newtonian dynamics at small accelerations (proportional to the square root of the cosmological constant). For this purpose, we use the framework of the precanonical quantization of gravity that was put forward in earlier 
 papers \cite{ikm1,ikm2,ikm3,ikm4} 
 and, more specifically, the precanonical quantization of tetrad gravity \cite{ikv1,ikv2,ikv3,ikv4,ikv5}. 
 This contribution further develops and corrects our earlier consideration of a possible quantum gravitational origin of MOND \cite{kk}.

Precanonical quantization of gravity uses a Hamiltonian-like formulation of multiple integral variational problems which is different from the standard canonical Hamiltonian formulation and requires no space+time decomposition. It is known as the De Donder-Weyl (DDW)  theory \cite{dedonder,kastrup}.  This formulation treats all spacetime dimensions on an equal footing and, therefore,  is not restricted to fields on globally hyperbolic spacetimes. To turn this formulation into a framework of a quantization of fields and gravity,  
 %it was necessary to find 
a proper generalization of Poisson brackets and Dirac brackets for singular DDW systems 
  %was found in earlier papers by one of us that  %%********************************
was found in \cite{ik5,mybr1,mybr2,mybr3,mydirac} 
that led to the brackets defined on differential forms and to the corresponding Poisson-Gerstenhaber algebra of dynamical variables 
represented by %them 
the differential forms. %$^{18-21}$  %\cite{ik5,mybr1,mybr2,mybr3}
The quantization of fields based on those structures is called the precanonical quantization. 
 %The precanonical quantization of scalar fields $^{17-21}$ leads to 
 It results in a hypercomplex generalization of quantum mechanics to field theory \cite{ik5e,ik2,ik3,ik4,ik5}
 where quantum fields are described in terms of 
a Clifford-algebra-valued  precanonical wave function on the bundle of field variables over spacetime and 
Clifford-algebra-valued operators whose representations are obtained from the Dirac quantization of the Heisenberg-like subalgebra of the Poisson-Gerstenhaber algebra. Such quantization of differential forms naturally leads to a hypercomplex generalization of quantum theory to  field theory, which is different from the picture that results from canonical quantization, which leads to an infinite-dimensional generalization of quantum mechanics to field theory. It was demonstrated \cite{iks1,iks2}
 that the latter can, in fact, 
be derived from the former using the $3+1$ decomposition and the 
de-quantization of Clifford algebra elements  at the quantum level back into differential forms at the classical level, namely,     
\beq \label{dequ}
\frac{1}{\ka}\gamma_\mu \mapsto \upsilon_\mu , 
%\der_\mu \inn (dx^0\we dx^1\we dx^2 \we dx^3) , 
\eeq 
where $\upsilon_\mu := \der_\mu \inn (dx^0\we dx^1\we dx^2 \we dx^3)$ and 
 $\ka$ is an ultraviolet parameter of the dimension of the inverse spatial volume. 
This parameter is introduced in precanonical quantization when the dual basis of differential forms  $\upsilon_\mu$ is represented by 
dimensionless elements of Clifford algebra.  The standard functional Schr\"odinger representation in quantum field theory 
emerges \cite{iks1,iks2} from the precanonical quantization in the case of infinitesimal $1/\ka$. 
It was demonstrated for scalar fields in flat \cite{iks1,iks2} and curved \cite{iksc1,iksc2,iksc3}  spacetimes, 
as well as for quantum Yang-Mills fields \cite{iky1,iky3}. 
Moreover, the result of the study of the spectrum of the DDW Hamiltonian operator of the SU(2) quantum Yang-Mills theory \cite{my-ymmg} 
suggests that the scale of physical $\ka$ is related to the hadronic scale of the mass gap in the nonabelian gauge theories of the Standard Model. This observation is crucial for the estimation of the cosmological constant and the Milgromian acceleration 
\cite{kk,my-mink,iktp1,iktp2}. 

%In this paper

We proceed as follows. In section 2, we outline the results of the precanonical quantization of the tetrad general relativity. 
It leads to the picture of quantum gravity as a spin connection foam described in terms of the wave function on the bundle of spin connection coefficients over spacetime. In section 3, we discuss the precanonical wave function of the quantum analogue of Minkowski spacetime. This is where the invariant acceleration scale $a_*$ emerges due to quantum fluctuations of spin connection. 
Then, in section 4, we discuss a modification of Newtonian dynamics by the quantum fluctuations of spin connection in the non-relativistic approximation. %of the spin connection foam.  
This allows us to explain the origin of the Milgromian MOND in section 5 and derive the theoretically motivated interpolating function for the case of a point mass. In section 6, we estimate the numerical value of parameters appearing in the theory and then discuss the observable effects of the spin connection foam and the respectively modified Newtonian dynamics in the Solar system, galaxies and larger structures in the Universe.

\section{Precanonical quantum gravity and spin connection foam}

 The precanonical quantization of pure tetrad gravity  \cite{ikv1,ikv2,ikv3,ikv4} 
%\cite{ikv1,ikv2,ikv3,ikv4} 
 starts from the following Palatini Lagrangian density for general relativity in tetrad variables:  
\beq \label{lagra}
{\mathfrak L}= \frac{1}{8\pi G} \left( {\mathfrak e} e^{\alpha}_I e^{\beta}_J \left(\der_{[\alpha} \omega_{\beta ]}^{IJ} +\omega_{[\alpha}^{IK}\omega_{\beta ] K}{}^J \right)  %\frac{1}{8\pi G} 
 - \Lambda {\mathfrak e}\right)  , 
\eeq
where  the spin connection coefficients 
$\omega_\alpha^{IK}$ and the tetrad components $e^{\alpha}_I$
are the independent field variables, $\Lambda$ is the cosmological constant,  and ${\mathfrak e} = \det(e^I_\mu)$.

According to the procedure of the DDW Hamiltonian formulation, we define the polymomenta of the field variables $e$ and $\omega$, which leads to the following primary constraints in the DDW Hamiltonian formulation: 
\begin{align} \label{constr}
\begin{split}
{\mathfrak p}{}^\alpha_{e^I_\beta} 
= \; \frac{\der {\mathfrak L} }{\der\, \der_\alpha e^I_\beta} \; \approx 0, \quad 
 {\mathfrak p}{}^\alpha_{\omega_\beta^{IJ}} &=\frac{\der {\mathfrak L} }{\der\, \der_\alpha{\omega_\beta^{IJ}}} \approx \frac{1}{8\pi G}
{\mathfrak e} e^{[\alpha}_Ie^{\beta ]}_{J },  
\end{split}
\end{align}
and the DDW Hamiltonian density on the surface of constraints (\ref{constr}):   
 \beq  \label{eham}
 %{\mathfrak H} 
 \mathfrak{e} H = {\mathfrak p}{}^\mu_{\omega_\alpha^{IJ}}\der_\mu \omega_\alpha^{IJ} + {\mathfrak p}{}^\mu_{e_\alpha^I} 
 \der_\mu e_\alpha^I - {\mathfrak L} 
 %+ \lambda ({\mathfrak p}{}_\omega - {\mathfrak e} e\wedge e) +\mu {\mathfrak p}{}_e 
 \approx -{\mathfrak p}{}^\alpha_{\omega_\beta^{IJ}}\omega_\alpha^{IK}\omega_{\beta K }{}^J 
 + \frac{1}{8\pi G}\Lambda {\mathfrak e} . %=: \mathfrak{e} H . 
 \eeq
The analysis of these constraints and the calculation of the corresponding generalized Dirac brackets of forms \cite{mydirac}  %$^{21}$  
 results in 
 %%very simple expressions such as ($\upsilon_{\alpha} := \der_\alpha \inn dx^0\we dx^1\we dx^2 \we dx^{3}$)
 \beqa 
  \label{dbr11c}
 {}&\pbr{{\mathfrak p}^\alpha_e}{e' \upsilon_{\alpha'}}{\!}^D=0, \nn \\
{}&\pbr{p^\alpha_\omega}{\omega'\upsilon_\beta}{\!}^D
%=\pbr{p^\alpha_\omega}{\omega'\upsilon_\beta} 
= \delta^\alpha_\beta \delta_\omega^{\omega'}, 
\label{dbr12c} 
 \\
{}&\!\!\hspace*{-10pt}\pbr{{\mathfrak p}^\alpha_e }{ {\mathfrak p}_\omega \upsilon_{\alpha'}}{\!}^D %\!=\!0\!=\! 
\!=\pbr{{\mathfrak p}^\alpha_e }{\omega \upsilon_{\alpha'}}{\!}^D 
\!=\pbr{{\mathfrak p}^\alpha_\omega}{e' \upsilon_{\alpha'}}{\!}^D \!=0 ,  \,\label{dbr13c} \nn
\eeqa 
and, similarly, in  simple brackets for other pairs of $3$- and $0$- forms of field variables 
$(e,\omega)$ and their polymomenta densities $({\mathfrak p}^\alpha_e, {\mathfrak p}^\alpha_\omega)$. 
 %where  $\upsilon_{\alpha} := \der_\alpha \inn dx^0\we dx^1\we dx^2 \we dx^{3}$. 

Quantization of form- and density-valued variables is based on the generalized  Dirac's quantization rule 
\beq
[\hat{F}, \hat{G}] = - i\hbar \widehat{\mathfrak{e} \pbr{F}{G}{}^D} , 
\eeq 
and it leads to the representation of the operators of the tetrad components
\beq
\hat{e}{}^\beta_I   = -  8\pi i G \hbar\ka   \ugamma{}^{J}\frac{\der}{\der \omega_{\beta}^{IJ}} ,\label{ebiop} 
\eeq
and the operator of the metric tensor 
\beq
\hat{g}{}^{\mu\nu} = - 64 \pi^2 G^2 \hbar^2\ka^2 \eta^{IK} \eta^{JL} \frac{\der^2}{\der {\omega_\mu^{IJ}} \der {\omega_\nu^{KL}}} ,   \label{gop}
\eeq
where 
\beq 
\ugamma^I \ugamma^J + \ugamma^J \ugamma^I = 2 \eta^{IJ}  , 
\eeq
and the parameter $\ka$ of the dimension of the inverse spatial volume appears on dimensional grounds as a general feature of the precanonical quantization.

The precanonical analog of the Schr\"odinger equation reads 
\beq  \label{pseq}
({i\hbar\varkappa%\slashed 
\not\!
\hat{\nabla} - \hat{H}}) \Psi =0 ,
\eeq
where $\hat{H}$ is the operator 
of the DDW Hamiltonian function 
\beq 
 \hat{H} = 8\pi G \hbar^2\ka^2 \
 \ugamma^{IJ}  
\omega_\alpha^{KM}\omega_{\beta M}{}^L \frac{\der}{\der \omega_{\beta}^{KL}} \frac{\der}{\der \omega_{\alpha}^{IJ}}
+ \frac{1}{8\pi G} \Lambda , \label{hDDWoper} 
\eeq 
and 
\beq 
\hat{\not\hspace*{-0.2em}\nabla} = 
- 8\pi i G \hbar\ka \ugamma{}^{IJ}\frac{\der}{\der \omega_{\mu}^{IJ}} 
\left(\der_\mu +  \frac{1}{4}\, \omega_{\mu KL}\ugamma{}^{KL} \stackrel{\leftrightarrow}{\vee}\right)         %\!\veelr) 
\label{nablaoper}
\eeq 
is the Dirac operator in which the curved space-time Dirac matrices $\gamma^\mu = e^\mu_I \ugamma^I$ 
render into differential operators because the tetrad components are differential operators (\ref{ebiop}). 
The spin connection term in (\ref{nablaoper}) acts on the Clifford-algebra-valued wave function  
$\Psi (\omega, x)$ by the commutator product 
\beq \label{cprod}
{\ugamma}{}^{IJ} \stackrel{\leftrightarrow}{\vee} \Psi 
 %:= \frac12\left( {\ugamma}^{bc} \vee \Psi - \Psi \vee {\ugamma}^{bc} \right) 
  = \frac12 \left[\ \ugamma{}^{IJ}, \Psi\ \right] . 
\eeq

%Using (\ref{hDDWoper}) and (\ref{nablaoper}), 
In the explicit form, the precanonical Schr\"odinger equation for quantum gravity %, eq. (\ref{pseq}), 
 %takes the form 
 reads 
\beq  \label{psegrav}
\ugamma{}^{IJ}  \frac{\der}{\der \omega_{\mu}^{IJ}}   %\veelr
   \Big ( \der_\mu + \frac{1}{4} \omega_{\mu}^{KL}\ugamma{}_{KL}\!\! \stackrel{{\leftrightarrow}}{\vee}  
 -\,
\omega_{\mu M}^{K}\omega_{\beta}^{ML}  \frac{\der}{\der \omega_{\beta}^{KL}} \Big)  %\veelr
 \Psi (\omega,x)     
 -  \lambda \Psi (\omega,x) = 0 ,  
 \eeq
where the dimensionless quantity 
\beq \label{lam}
\lambda = \frac{\Lambda}{ (8\pi G \hbar\varkappa)^2}    
\eeq 
contains both the fundamental constants $G$ and $\hbar$ and the parameter $\ka$ introduced by the precanonical quantization.

%is a dimensionless combination of the fundamental constants of the theory, which depends on the operator ordering of 
%$\omega$ and $\frac{\der}{\der \omega}$. 

% The operators act on Clifford-algebra-valued precanonical wave functions on the configuration bundle of spin connection variables over %spacetime variables, $\Psi(\omega,x)$, whose 
 
 The invariant scalar product of Clifford-algebra-valued precanonical wave functions has the form 
\beq \label{scprod}
\left\langle \Phi | \Psi \right\rangle 
=  %\Tr \int \Phib \, \hat{[d\omega]}_{} \Psi , 
\mathrm{Tr} \! \int \prod_{\mu, I,J} d \omega_\mu^{IJ} \, \left( \Phib \, \hat{\mathfrak e}{}^{-6} \Psi \right ),
\eeq
where $\Phib =\ugamma{}^0\Phi^\dagger\ugamma{}^0$    % = \Psi^*{}^r$,  
and %the operator-valued invariant integration measure on the 
  %$\frac12 n^2(n-1)$-dimensional 
  %$24$-dimensional space of spin connection coefficients 
the  operator of ${\mathfrak e}{}^{-6} = {\det(e^\mu_I)}{}^6$  
 can be also understood 
 as the operator $\widehat{\det(g^{\mu\nu})}{}^3$ which can be constructed from the representation in (\ref{gop}). 

The physical picture of quantum gravity according to the precanonical quantization is what we call the spin connection foam. 
The latter is characterized   by the Clifford-algebra-valued 1-point amplitudes $\Psi(\omega, x)$ and the 
2-point amplitudes $\left< \omega, x | \omega', x'\right>$ that are Green's functions for equation (\ref{psegrav}). 
They describe the probability distribution and correlations of quantum fluctuations of spin connection at different spacetime points.   

%The normalizability of %$\Psi(\omega)$: precanonical wave functions: $\left\langle \Psi | \Psi \right\rangle < \infty$, 
%leads to the vanishing contribution of the large curvatures $R=d\omega + \omega\we\omega$   
%to the probabilistic measure defined by the norm, 
%%**** and thus  leads to a quantum-gravitational curvature  singularity avoidance. 
% and that ensures the quantum-gravitational avoidance of curvature  singularities by the precanonical wave function. 

%The evolution of matter and radiation on the background of quantum gravitational fluctuations 
% whose statistics and correlations are predicted by (\ref{psegrav}) may lead to observable  consequences for the distribution of 
% matter and radiation  at large cosmological scales. 

%\newpage 

%OLD

\section{Quantum Minkowski spacetime}
  
Let us construct the spin connection foam corresponding to the Minkowski spacetime \cite{my-mink}. 
The Minkowski spacetime in Cartesian coordinates can be characterized by the vanishing spin connection coefficients 
$\omega_\mu^{IJ} = 0$. 
%\beq \label{om}
% \omega_\mu^{IJ} = 0.
% \eeq
In this case, precanonical Schr\"odinger equation (\ref{psegrav}) with $\Lambda = 0$ assumes a simple form 
\beq  \label{pseqm}
\ugamma^{IJ}\frac{\der}{\der {\omega_\alpha^{IJ}}} \frac{\der}{\der x^\alpha} \Psi (\omega, x)= 0 . 
\eeq
The square of this equation  
\beq\label{pseqm2}
\eta^{IK} \eta^{JL}\frac{\der}{\der x^\alpha} \frac{\der}{\der x^\beta}  \frac{\der}{\der \omega_\alpha^{IJ}} \frac{\der}{\der \omega_\beta^{KL}} \Psi = 0  
\eeq
decouples different components of $\Psi$ so that we can assume here that $\Psi$ is a scalar. 
We combine Eq. (\ref{pseqm2}) with the requirement of the correspondence of the average of the metric operator 
$\hat{g}^{\mu\nu}$  given by (\ref{gop}) 
with the Minkowski metric $\eta^{\mu\nu}$ on average:  
\beq\label{mi9}
\langle \hat{g}^{\mu\nu}\rangle (x) := {\mathrm{Tr}}\! \int\! d^{24}\omega %\delta^{24}(\omega) 
\left( \Psib(\omega,x) 
\hat{\mathfrak{e}}{}^{-6}\hat{g}^{\mu\nu} \Psi (\omega,x) \right) = \eta^{\mu\nu} .  
\eeq 
Then equation (\ref{pseqm2}) 
implies that the modes of the precanonical wave function on the spin connection bundle satisfy 
\beq \label{helm}
\eta^{IK} \eta^{JL} \frac{\der}{\der \omega_\alpha^{IJ}} \frac{\der}{\der \omega_\beta^{KL}} \Psi 
+  \frac{1}{(8\pi G\hbar \ka)^2}\eta^{\alpha\beta} \Psi = 0 
\eeq
and 
\beq \label{dalambert}
\eta^{\mu\nu} \frac{\der}{\der x^\mu} \frac{\der}{\der x^\nu} \Psi = 0 . 
\eeq 
Thus the precanonical wave function of the Minkowski spacetime has light-like modes propagating on the base of the spin connection bundle, as described by the d'Alembert equation in (\ref{dalambert}), 
and massive modes in the fibers, the space of spin connection coefficients, 
 as described by the wave equations in (\ref{helm}). 
 
 Note that, according to (\ref{helm}), the range of the massive modes in the space of spin connection coefficients 
 \beq \label{astar}
a_* = 8\pi G \hbar\ka 
\eeq 
 defines an invariant scale of accelerations (in the units in which the speed of light is set equal to unity).  
%(in the units with $c=1$)
%(in the units in which $c=1$)
At this scale, the classical notion of inertial frames is violated by quantum fluctuations in the spin connection foam. 
As a consequence, classical dynamical laws can be modified accordingly, as we will show in the next section. 
 %when the accelerations due to the external fields $a \lesssim a_*$. 
 %at small accelerations of the order of or smaller than $a_*$. 
 %$a \lesssim a_*$.
 
 Note also that (\ref{lam}) and (\ref{astar}) imply 
 \beq \label{astarla}
 a_* = \sqrt{\Lambda/\lambda} , 
 \eeq 
 where the constant $\lambda$ is determined by a proper ordering of operators in (\ref{psegrav}). 
This relation between the acceleration scale $a_*$ and the square root of the cosmological constant 
resembles the relation between the Milgromian acceleration scale $a_0$ and the square root of the cosmological constant  
in the modified Newtonian dynamics (MOND) which was proposed by Milgrom  \cite{mond83,mond,mond6,mond-th}  
%$^{38-41}$ 
as an explanation of flat galactic rotation curves 
without dark matter \cite{mond5}.  The emergence of the Milgromian MOND from the averaged dynamics of nonrelativistic test particles 
in the nonrelativistic approximation of the spin connection foam will be discussed in section 5.

%\section{The cosmological constant} 

%From (\ref{lam}),  
%\beq \label{Lala}
%{\Lambda} = \lambda{ (8\pi G \hbar\varkappa)^2}, 
%\eeq
%where the constant $\lambda$ depends on the ordering of operators in (\ref{psegrav}). The ordering is fixed by requiring 
%the terms in (\ref{psegrav}) which do not contain the spacetime derivatives $\der_\mu$ to be symmetric operators on the space 
%of Clifford-valued wave functions equipped with the scalar product (\ref{scprod}).  
 %\cite{kk23}. 
% The contribution from the Weyl ordered spin connection operator $\frac{1}{4}  \ugamma{}^{IJ}  \der_{\omega_{\mu}^{IJ}}   %\veelr
%  \omega_{\mu}^{KL}\ugamma{}_{KL}\!\! \stackrel{{\leftrightarrow}}{\vee}$  is 
%\beq 
%\lambda = -  \frac{1}{16}\,  \ugamma{}^{IJ} \ugamma{}_{KL}  \left[ \der_{\omega_{\mu}^{IJ}} ,   %\veelr
%   \omega_{\mu}^{KL}\right]   = 3 , %1/4  
%\eeq    
%(it differs by a factor $1/16$ from a rough estimation in \cite{my-mink}). 
%compare with the estimation in ....
 %\cite{my-mink}. 

\section{Quantum gravitational modification of the Newtonian dynamics (qMOND)}

Here we will restrict ourselves to the motion of non-relativistic test particles. The geodesic equation 
in the non-relativistic approximation reads 
\beq  \label{mo3}
\ddot{x}{}^i + \Gamma^i_{00}= 0 . 
\eeq
In this approximation, the components of the Christoffel symbol $\Gamma^i_{00}$ are equal to the components of spin connection 
$\omega^{i 0}_0$ to be denoted by $\omega^{i}$. For the gravitational field of a point mass $M$ at the center of coordinates,  
$\omega^{i} = GMx^i/r^3$, where $r^2 = x_i x^i$. We assume that this system is immersed into a non-relativistic limit 
of the quantum Minkowski space that was described in section 3. The non-relativistic limit means 
that the only fluctuating components 
of spin connection are $\tilde{\omega}{}^i$ 
  with $\langle \tilde{\omega}{}^i \rangle = 0$ 
%$\omega^{i0}_0$ 
and their statistics is described by the solutions of the $00$-component of equations (\ref{helm}): 
\beq\label{mo20}
\eta^{ij} \der_{\tilde{\omega}{}^i} \der_{\tilde{\omega}{}^j} \Psi  
= - \frac{1}{(8\pi G \hbar\ka)^2} \eta^{00} \Psi .    %\eta^{00} \Psi  ,
\eeq 
The ground state solution of this  modified Helmholtz equation has the form of the Yukawa potential
  %%whose real-valued ground state (Yukawa) solution and its normalization are %read  
\beq \label{mo22}
\Psi (\tilde{\omega}{}^i) 
 = %\mathcal{N}{}^{-1/2}   
\frac{1}{\sqrt{16\pi^2 G \hbar\ka}\ \omega} %% %% 8 > 16 corrected 2026
 %\ \omega^{-1} 
\ e^{-\omega/(8\pi G \hbar\ka) } , %\quad \langle \Psib |\Psi \rangle = 1 , 
%\quad \omega : = \sqrt{\tilde{\omega}{}^i \tilde{\omega}{}_i} , 
\eeq 
 where $\omega  = \sqrt{(\tilde{\omega}{}^i){}^2} \ge 0$ and the normalization factor 
 ${1}/{\sqrt{16\pi^2 G \hbar\ka}}$ %% %% 8 > 16 corrected 2026
 is derived from the following requirement 
\beq
\langle \Psi |\Psi \rangle = \int\! d {\tilde{\omega}{}^1} d {\tilde{\omega}{}^2} d{\tilde{\omega}{}^3} \, 
\Psi(\tilde{\omega}{}^i)^2  = 1  .
\eeq
Note that this ground state is represented by a real valued spherically symmetric wave function. 
For the relevant expectation values we obtain 
\begin{align}
\label{avom0}
\langle \tilde{\omega}{}^i \rangle 
&= \int\! d {\tilde{\omega}{}^1} d {\tilde{\omega}{}^2} d{\tilde{\omega}{}^3} \,  \tilde{\omega}{}^i \, \Psi^2
= 0 , \\ 
\label{avom2}
\langle \tilde{\omega}{}^i \tilde{\omega}{}_i \rangle 
 &  %=: \bar{a}{}^2 
= \int\! d {\tilde{\omega}{}^1} d {\tilde{\omega}{}^2} d{\tilde{\omega}{}^3} \, 
\Psi \omega^2 \Psi   
= 4\pi  \int_0^\infty\! d {\omega} \, \omega^4 \Psi^2 %%2pi > 4pi corr 2026
 %= ... 
=\frac{1}{2} (8\pi G \hbar\ka)^2 
= \frac{1}{2} a_*^2 , \\ 
\label{avom1}
\langle \sqrt{\tilde{\omega}{}^i \tilde{\omega}{}_i}\, \rangle 
&= 4\pi  \int_0^\infty\! d {\omega} \, \omega^3 \Psi^2 = 4\pi G \hbar\ka = \half a_*  . %%2pi > 4pi corr 2026
\end{align}
%
%and, by denoting $\langle \ddot{x}{}^i\ddot{x}_i \rangle %=(\ddot{x}{}^i)^2 =: a^2$, we obtain ,.,...,

Now, the movement of a test particle in the gravitational field of the point mass $M$ immersed in the nonrelativistic approximation of the spin connection foam can be described by the equation  
\beq  \label{mo3a}
\ddot{x}{}^i 
%\Gamma^i_{00}= %\omega^i =
%\omega^{i 0}_0 = GM \frac{x^i}{r^3} + \tilde{\omega}{}^i . %\quad \langle \tilde{\omega}{}^i \rangle = 0.
%%- \Gamma^i_{00}= %\omega^i =
%%- \omega^{i 0}_0 = -
 + GM \frac{x^i}{r^3} + \tilde{\omega}{}^i = 0  . 
 %, \quad \langle \tilde{\omega}{}^i \rangle = 0 . 
\eeq
Since $\langle \tilde{\omega}{}^i \rangle = 0$, the average of this equation just reproduces the classical equation of motion in (\ref{mo3}). However, the nonvanishing variance 
$\langle \tilde{\omega}{}^i \tilde{\omega}{}_i\rangle = \langle \tilde{\omega}{}^2\rangle  \neq 0$ leads to non-trivial consequences of the average of the square of (\ref{mo3a}): 
\beq \label{square}
\left\langle \ddot{x}{}^2 + \frac{G^2M^2}{r^4} + \tilde{\omega}{}^2 + 2GM \frac{\ddot{x}{}^ix_i}{r^3} 
+ 2GM \frac{{x}{}^i\tilde{\omega}{}_i}{r^3} + 2 \ddot{x}{}^i \tilde{\omega}{}_i \right\rangle   = 0 . 
\eeq
Using (\ref{mo3a}) again, we can rewrite (\ref{square}) in the form 
%\beq \label{square1}
%\left\langle \ddot{x}{}^2 + \frac{G^2M^2}{r^4} + \tilde{\omega}{}^2 
%%%+ 2GM \frac{\ddot{x}{}^ix_i}{r^3} 
%+ 2GM \frac{\ddot{x}{}^i\tilde{\omega}{}_i}{r^3} 
%%% + 2 \ddot{x}{}^i \tilde{\omega}{}_i 
%- 2\ddot{x}{}^2
% \right\rangle   = 0 
%\eeq
%or 
\beq \label{square2}
\left\langle \ddot{x}{}^2 - \frac{G^2M^2}{r^4} - \tilde{\omega}{}^2 
%+ 2GM \frac{\ddot{x}{}^ix_i}{r^3} 
- 2GM \frac{{x}{}^i\tilde{\omega}{}_i}{r^3} 
% + 2 \ddot{x}{}^i \tilde{\omega}{}_i 
%%- 2\ddot{x}{}^2
 \right\rangle   = 0 . 
\eeq
Within our assumption underlying equation (\ref{mo3a}), %%?
 that the Newtonian gravitational system of the point mass and the test particle 
 is immersed in the quantum spacetime with fluctuating spin connections, it is natural to assume that 
\beq \label{zerocorr}
\left \langle 
\frac{{x}{}^i\tilde{\omega}{}_i}{r^3} 
% + 2 \ddot{x}{}^i \tilde{\omega}{}_i 
%%- 2\ddot{x}{}^2
 \right\rangle   = 0 .
\eeq
In fact, if the averaging is over quantum fluctuations of spin connections, it is a direct consequence of $\langle \omega^i \rangle = 0$. Therefore, by denoting 
%$\ddot{x}^2 = a^2$ 
$a = |\ddot{x}| = \sqrt{\ddot{x}^2}$  and $\abar{}^2 =  \langle \tilde{\omega}{}^2 \rangle$, 
we obtain the following  modified  Newton's law of universal gravitation for a point mass 
  which we propose to call quantum-gravitationally  modified Newtonian dynamics or qMOND 
\beq \label{agener}
a = %\sqrt{\frac{G^2M^2}{r^4} + \tilde{\omega}{}^2 } 
\sqrt{\frac{G^2M^2}{r^4} + \abar{}^2} . 
\eeq
The right hand side can be considered as the radial component of the negative gradient of the following 
 qMOND generalization 
 %of the corrected 
 of the Newtonian gravitational potential of a point mass which 
  %in qMOND 
  takes into account non-relativistic quantum fluctuations of spin connection:\footnote{After this paper was submitted in December 2024, a simpler hypergeometric expression of this potential was found in our subsequent publications.\cite{epl25,dice24,mpla25}}  
\beq \label{phir}
\Phi(r) =  - \frac{GM}{r} \left( 1+ \frac{\abar^2 r^4}{G^2M^2} \right)^{3/2}  
{}_2F_1 \left(1,\frac54;\frac34; - \frac{\abar^2 r^4}{G^2M^2}\right) , 
\eeq 
where $_2F_1 (a,b;c;z)$ is the standard Gauss hypergeometric function and where the Planck's constant $\hbar$ is hidden in $\abar$ 
as is (\ref{astar}) and  (\ref{avom2}). 
For small $r$, the Newton's law $a = {GM}/{r^2}$ for the absolute values  is reproduced, 
and for large  $r$,  
%\rightarrow \infty$, 
$a = \abar$,  %.... interpretation?? sign $\abar$?? .......\\ 
which corresponds to the linear potential $\Phi (r) = \abar r$ that emerges here as the ``anti-screening"  
effect of fluctuations of spin connection. 
Interestingly, this asymptotic behavior also appears in the Cornell potential \cite{cornell1} postulated 
in the context of the quark confinement problem, in which the parameter $\abar$ is replaced by a hadronic string tension parameter,  
 %%%anti-screening!!
 in Grumiller's  model of ``gravity at large distances" \cite{grum} with $\Lambda = 0$ and $\abar$ identified with the Rindler acceleration, 
 and in spherically symmetric solutions of conformal Weyl gravity  \cite{mannheim,conf-modesto} 
 which are able to describe galactic rotation curves by going beyond Einstein's General Relativity instead of postulating dark matter. 
 
 %which are able to describe 

 %%+ https://arxiv.org/pdf/1107.2373

The vector form of the qMOND generalizaton of Newton's law can be written in the form 
\beq \label{vector}
\vec{a} =  - \vec{\nabla} \Phi . 
\eeq
For a spherically symmetric potential $\Phi(r)$ it is tantamount to 
%%has the same effect as 
multiplying the scalar equations by the unit vector $\vec{r}/r$.

When classical gravitation dominates over quantum fluctuations of spin connection, i.e. 
$\frac{GM}{r^2} \gg \abar{}$, we obtain the following correction 
\beq \label{appr1}
a \approx \frac{GM}{r^2} + \frac{\abar{}^2 r^2}{2 GM} .
\eeq

When quantum fluctuations of spin connection dominate over the classical gravitation, i.e. $\abar \gg \frac{GM}{r^2}$, 
we obtain 
\beq \label{appr2}
a \approx \abar + \frac{G^2M^2}{2 \abar{} r^4} 
- \frac{1}{8} \frac{G^4 M^4}{\abar^3 r^8}  , 
\eeq
  where the last two terms are similar to the force that can be derived  from the  ``7-3 Lennard-Jones potential". 
  % first introduced by Gustav Mie \cite{mie1903}. 

It is interesting to note that, in both cases, the geometric mean of the gravitational radius of the mass $M, r_S=2GM$, and the cosmological radius $r_\Lambda \sim \abar{}^{-1}$, namely,   $r_* = \sqrt{r_Sr_\Lambda}$, is the border between two different physical regimes.

\section{The origin of MOND}

Due to the omnipresent quantum fluctuations of spin connection, 
  %the physics can not be formulated in an inertial system. 
  %Instead, we assume that the accelerations are being measured with respect to
  classical accelerations are being measured with respect to the ``mean field" acceleration $\abar$. 
In this reference system, the standard Newton's law (for the absolute values) reads 
\beq \label{defo}
a - \abar = \frac{GM}{r^2} . %\quad (??? + or -) \quad \mathtt{Note\; a\; flip\; of\; sign!}
\eeq 
 We can take this equation as a redefinition of the right hand side in terms of the left hand side, 
which is based on non-relativistic experiments at normal accelerations $a \gg \abar$,  and then try to extrapolate its validity 
(i.e. the definition of quantities which is based on the domain of classical Newtonian physics) to arbitrary accelerations. 
Then, equation (\ref{appr2}) without the last $O(r^{-8})$ term  can be rewriten in terms of the acceleration with respect to 
the mean field background 
$g = a- \abar$ as follows 
\beq \label{milga} 
 \frac{GM}{r^2} = \frac{g^2}{2 \abar{}} . 
\eeq
Therefore, in the regime of small accelerations $g$, we reproduce the equation of Milgrom's Modified Newtonian Dynamics (MOND) 
in the deep-MOND regime 
\beq \label{milgb}
\frac{GM}{r^2} = \frac{g^2}{g_0}   , 
\eeq \label{go}
where the Milgromian acceleration 
\beq 
g_0 = 2 \abar .
\eeq 

Moreover, if 
 %we do the same with the 
 we apply the same procedure to the 
 qMOND law in equation (\ref{agener}), we obtain 
\beq  \label{mmondab}
\frac{GM}{r^2} =  \sqrt{g^2 + \abar{}^2} - \abar  . 
\eeq

According to MOND, the modified dynamics for arbitrary accelerations has the form \cite{mond83,mond}
\beq \label{mmond}
\frac{GM}{r^2} = \mu\left( \frac{g}{g_0}\right)  g , 
\eeq
where $\mu(x)$ is an interpolating function such that $\mu(x\rightarrow 0) \rightarrow x$ and 
$\mu(x\rightarrow \infty) \rightarrow 1$, so that the Newtonian dynamics is reproduced at accelerations 
$g\gg g_0$ and the deep-MOND regime (which guarantees the flat asymptotic of the rotation curves of galaxies) 
is reproduced at $g \ll g_0$.   The choice of the interpolating function is usually based on the consideration 
simplicity and phenomenology.  
 %refs 
For example, the choice of the ``simple interpolating function"  $\mu_{\mathtt{simple}}(x) = x/(1+x)$ has been proven to be phenomenologically viable \cite{}.  
However, there are no theoretical arguments in favor of a specific choice of the interpolating function which may be considered 
a major weakness of the theory of MOND in its present form.

By comparing (\ref{mmondab}) with (\ref{mmond}) we obtain 
\beq  \label{mux}
\mu(x) = \frac{1}{2x} \left(\sqrt{4x^2+1} - 1    \right) , 
\eeq
with $x= {g}/{g_0}$ and $g_0 = 2\abar$. It is easy to see that this interpolating function satisfies the required 
asymptotics for $x\rightarrow 0$ and $x \rightarrow \infty$. We, thus,  derive the theoretical interpolating function $\mu(x)$ from the first principles of precanonical quantum gravity and the approximations %underlying our derivation of MOND.  
 %which have allowed us to derive MOND from it. 
 that allow us to derive MOND as a manifestation of the spin connection foam. 
 %It is interesting to note that 
 This interpolating function  very closely matches $\mu_{\mathtt{simple}}(x)$ with the 
maximal deviation of around $12\%$ near $x\approx 1.3$ and less than $2\%$ in the deep-MOND regime, thus making it phenomenologically viable \cite{mond5}.  
% Quite unexpetedly, 
 Let us note that, surprisingly, the same interpolating function as in (\ref{mux}) was also obtained from the considerations \cite{milg99,pikh10,klink12} which are based on 
% the  hypothetical first principles which are different from  those underlying this paper. 
different underlying 
 %first principles 
 hypotheses than this paper.

%%cf . Klinkhamer, F.R.: Entropic-gravity derivation of MOND. arXiv:1201.4160 [gr-qc]
%%P. V. Pikhitsa, MOND reveals the thermodynamics of gravity, arXiv:1010.0318.
%%M. Milgrom, “The modified dynamics as a vacuum effect,” Phys. Lett. A 253, 273 (1999), arXiv:astro-ph/9805346.

\section{Discussion and outlook } 

The acceleration $a_*$ and the cosmological constant $\Lambda$ depend both on the Planck scale of $G\hbar$ and the unknown scale of the 
parameter $\ka$. The study of the spectrum of the DDW Hamiltonian of quantum Yang-Mills theory \cite{my-ymmg}, that controls the mass spectrum of the theory,  has shown that the order of magnitude of the mass gap $\Delta m$ is related to the scale of $\ka$ as follows: 
\beq \label{kaest}
%\Delta m\! \sim\! (g^2\hbar^4\ka)^{1/3} , 
\varkappa \sim \frac{(\Delta m)^3}{\hbar^4 g_s^2} , 
\eeq
where $g_s$ is the gauge coupling constant in the classical YM Lagrangian. This formula is based on a rather rough 
spectral estimate %is rather rough 
and it does not take into account a factor that depends on the dimension of the gauge group.  Then the invariant scale of acceleration 
$a_*$ in (\ref{astar}) is given by 
\beq \label{astarn}
a_* \sim 8\pi G \frac{(\Delta m)^3}{\hbar^3 g_s^2} .
\eeq 
It is natural to identify the classical coupling constant $g_s$ with the QCD running constant at the momentum transfer $Q=0$. Although the perturbative result for the running coupling constant leads to the Landau pole at $Q=0$,  the non-perturbative consideration and the experimental data indicate  \cite{deur16,deur23,sanctis24}  that the infra-red limit of 
$g_s^2(Q) = 4\pi \alpha_s(Q)$ stabilizes to the value $g^2_s \approx 4\pi^2$ at $Q\rightarrow 0$.

The mass gap in QCD is usually expected to be in between the meson scale of $\sim$100MeV and the mass of the glueball candidates at a few GeV. If we assume that the mass gap corresponds to the mass of the lowest observable excitation in QCD at $\Delta m \sim 10^{- 1}\, \mathrm{GeV}$, 
from (\ref{astarn}) we obtain the numerical value 
\beq \label{astarnum}
a_* \sim 10^{-27}\, \mathrm{m}^{-1} . 
\eeq 
This is consistent with the value of the Milgromian acceleration  $g_0 = \sqrt{2}a_* = 0.12\, \mathrm{nm}/s^{2} $ or 
$\approx 10^{-27}\, \mathrm{m}^{-1} $ in geometrized units,  although the error of this estimation is several orders of magnitude 
since the accumulated uncertainties in the exact values of the quantities and the factor in (\ref{kaest}) are of several orders of magnitude. The same approximate consistency up to several orders of magnitude with the observed cosmological constant 
$\Lambda \approx 10^{-52}\, \mathrm{m}^{-2}$  follows from the relation between $a_*$ and $\Lambda$ in (\ref{astarla}) 
 with $\lambda=3$ for the Weyl ordered spin connection operator in (\ref{psegrav}) \cite{kk}. 
 
Thus the theory of precanonical quantization of General Relativity predicts a Universe  with a 
dark energy in the form of a small cosmological constant 
and the related fundamental  acceleration scale, both emerging from quantum fluctuations of the spin connection foam that 
also leads, 
as we will see below,  to the effects which are usually attributed to a hypothetical 
 %dark energy and 
dark matter. Based on the above, we assume  
 %We, therefore, assume 
 in what follows that the value of $\abar = g_0/2$ is obtained from the empirical value of the Milgromian acceleration $g_0$.

%Lambda 
% In QCD, $\Delta m \sim 10^{0\pm 1} \mathrm{GeV}$  and $g^2 \sim 10^0$ 
%.....%(see \cite{iktp2} and the references therein), and, therefore, 
%... by taking into account all the current 
%uncertainties in the values of $\Delta m$, the gauge coupling $g$, and the spectral estimate in 
    %\cite{my-ymmg}, g_s := 
%we conclude that $\ka \sim 10^{0 \pm 6} \mathrm{GeV}^3$. Consequently, from (\ref{astar}) and (\ref{Lala}), we obtain  
%\beq
%a_* \sim 10^{-23\pm 6} \mathrm{cm}^{-1} \quad \mathrm{and} \quad \Lambda \sim  10^{-46 \pm 12} \mathrm{cm}^{-2} .  
%\eeq 
%These values overlap with the values of the Milgromian acceleration 
 %$a_0 \approax 1.2 \times 10^{-10} \mathrm{m\cdots^{-2}} $  
 %$a_0 \approx 10^{-29} \mathrm{cm}^{-1}$   
%and the cosmological constant $${\Lambda} \approx 10^{-56} \mathrm{cm}^{-2}$, respectively. 
%$a_* = 8\pi G\hbar\ka = g_0/\sqrt{2} = 0.9 \times 10^{-10} \mathrm{m s}^{-2}$ 
%assuming the Milgromian acceleration $g_0 = 1.2 \times 10^{-10} \mathrm{m s}^{-2}$  or $0.12 \mathrm{nm s}^{-2}$

\bigskip

Then,  in the {\bf Solar system} with $GM_\odot \approx 1.5\, \mathrm{km}$ (in geometrized units),   
the correction in (\ref{appr1}) will be $1\%$ of the gravitational acceleration from the Sun at the distance  
\beq
r_M = \left( 0.01 \times \frac{2G^2M^2_\odot}{\abar^2 }\right)^{\frac14} \sim 3\times 10^3\, \mathrm{au} , 
\eeq
i.e. the deviation from the Newtonian dynamics may be detectable for objects from the 
inner edge of the \"Opik-Oort cloud. The deviation at the current location of Voyager 1 spacecraft (at 166\! $\mathrm{au}$ from the Sun)  is about $10^{-4}\%$. 

%Moreover, w
We have also shown \cite{kk25} that 
the effects of the spin connection foam  in the vicinity of the Earth orbit 
lead to the corrections of the order of $10^{-9} \%$
to the duration of the Earth year $T$ and the locations of the Lagrange points, both based on 
the following correction 
 %the $O(\abar^2) correction$ 
 to Kepler's  third law: 
\beq
%\omega^2
\left (\frac{2\pi}{T} \right )^2 = \frac{G(M_1 + M_2)}{R^3} + \abar^2\, %\frac{R}{2G} 
\frac{R(M_1 + M_2)}{2G M_1 M_2} + O (\abar^4) , 
\eeq
where 
$R$ is the semi-major axis of the Earth orbit, and $M_1$ and  $M_2$ are the masses of the Sun and Earth, respectively.

Moreover,  for the mass $M=1$\,kg with $GM\sim 10^{-27}$\,m  the condition  $GM/r^2 \gg \abar$ is satisfied for  $r \ll 1$\,m. 
At the distance  $r = 10$\,cm from the mass $M$ the correction to the Newtonian acceleration in (\ref{appr1}) is $\,\sim\!10^{-2} g_0 \approx 10^{-12}$\,m/s$^2$. When acting on the test mass of $1$\,mg it leads to the force of $10^{-18}$N. As the sub-attonewton sensitivity of force sensors is already reachable \cite{subatto1,subatto2},  and the masses and distances in our estimation here are compatible with the experimental setup in \cite{atto}, which has already achieved the attonewton sensitivity, our correction in (\ref{appr1}) 
can already be experimentally tested.  %in a lab. 

%According to the correction in (\ref{appr1}), the gravitational attraction of two mg particles separated by a distance of 1 mm 
%will experience almost 100 $\%$ increase in the force of their gravitational attraction with respect to the prediction of 
%the Newtonian gravity. 

\bigskip 
%For a galaxy of baryon mass $\mathfrak{M} \sim 10^{11} M_\odot$, $G \mathfrak{M} \approx 10^-2 ly$,  
%............. 

For the movement around point masses of the order of the mass $\mathfrak{M}$ 
of {\bf galaxies}, we can derive from (\ref{agener})  the formula for the orbital velocities $v$ such that $v^2/r = a$, 
\beq \label{vr}
v (r) = \left( \frac{G^2 \mathfrak{M} ^2}{r^2} + \abar^2 r^2  \right)^{1/4} . 
\eeq
This function has the minimum at
\beq \label{rm}
r_m = \sqrt{ \frac{G \mathfrak{M}}{\abar}} , 
\eeq
where the velocity 
\beq \label{vm}
v_m := v(r_m) = (2\abar G \mathfrak{M})^{1/4} .
\eeq
Near this point the rotation curve $v(r)$ is approximated by %a very flat parabola 
\beq
v(r) = (2\abar G \mathfrak{M})^{1/4} +  \frac{\abar^2}{v_m^3} (r - r_m)^2 + O ((r-r_m)^3) .
\eeq 
This is a very flat parabola around $r_m$ since the scale of $\abar$ is cosmological and the scale of $G \mathfrak{M}$ for most galaxies is below $1\, \mathrm{ly}$. The first term corresponds to the velocity of flat rotation curves according to MOND and the 
phenomenological Baryonic Tully-Fisher relation between the visible (baryonic) mass of a galaxy and the velocity in the flat part 
of its rotation curve. 

For example, for a galaxy of the baryonic mass $\mathfrak{M} \sim 10^{11} M_\odot$, 
$G \mathfrak{M} \approx 1.5 \times 10^{-2}\, \mathrm{ly}$,  $r_m \approx 5 \times 10^4\,  \mathrm{ly}$,  and 
$v_m \approx 0.65 \times 10^{-3}$ (i.e. $195$ km/s). With the error margin of $\pm 10\% $,  %this flat parabola 
the rotation velocity $v(r)$ in (\ref{vr}) 
can be approximated by a flat rotation curve $v(r) \approx 210$ km/s in the region between $30\, \mathrm{kly}$ and $90\, \mathrm{kly}$. 
It  is consistent with the observed flat rotation curves of such galaxies (like M31 or Milky Way \cite{m31-24,m31-06,mw-14}) considering the fact that the behavior of the rotation curves at $r<30\, \mathrm{kly}$ 
 %to a larger extent is 
 is predominantly determined  by the 
 %internal 
 mass distribution in the galactic disk \cite{mond5}, 
 which our approximation of the central point mass here completely ignores.    
 
 At larger distances $r>100\, \mathrm{kly}$ ($31\, \mathrm{kpc}$),  
 equation (\ref{vr}) predicts the linear growth of the rotation velocity at the scale of hundreds of kly and then,  at even larger distances, the $\sqrt{\abar r}$ growth, as dictated by the asymptotic linear behavior of the potential $\Phi(r)$ in (\ref{phir}). 
 The hierarchy of 
scales of different regimes which follows from  the formula in (\ref{vr}) for the rotation curve $v(r)$ around 
an isolated point mass $M$ and its potential $\Phi(r)$ in (\ref{phir}),  suggests  that it can play a key role in the dynamics of galactic clusters, superclusters, voids, and larger scale structures formation. Because the potential $\Phi(r)$ ensures deeper and longer range  potential wells,  it can account for even more of the missing mass than MOND at those scales, 
 as a MOND-like description of galaxy clusters  requires \cite{clusters},  
 and it can also explain a faster clumping of matter than expected in the standard $\Lambda$CDM and even MOND \cite{mcgaugh24} due to it allowing  for the earlier mass fluctuations to attract 
%grab 
%(attract, seize??) 
matter from the more distant regions of space due to the asymptotic linear grown of the potential $\Phi(r)$. Moreover, it may be interesting to check via numerical simulations if the one-dimensional nature of the linear potential up to a Gly scale, where the effects of the cosmological background will take over,  could actually promote the formation of quasi one-dimensional large scale structures like  the galaxy filaments in the cosmic web. 

%However, 
It should be noted that the linear and the square root growth of the rotation velocity at the distances above $100\, \mathrm{kly}$  from the gravitating center of $10^{11}$ solar masses, which follows from equation (\ref{vr}),  seems to contradict 
 observations of flat rotation curves at the Mly scales \cite{1mpc}, 
while those are described by MOND by design. 
However, we have shown in section 5 that the Milgromian MOND is obtained 
  %in the non-inertial reference frame 
  by  accounting for the non-inertial effects of the mean field of 
fluctuating accelerations in the quantum spin connection foam. This is why the observable effects of the interactions of very distant objects via the potential in (\ref{phir})  %are >> 
 can be 
 described by MOND using the accelerations 
$g, g_0$  and an interpolating function 
  like the one we have derived in (\ref{mux}). 
  %% or other phenomenological ones. 
 In more realistic situations, however, the interpolating function 
should also take into account both the multi-body interactions (mass distributions) and correlations between macroscopic massive bodies immersed in the quantum spin connection foam (as a field of fluctuating accelerations),  in accordance with the equivalence principle and  the two-point solutions of the precanonical Schr\"odinger equation in (\ref{pseq}). 
 The work on addressing those issues is in progress. 
%We are going to address these issues in a future work  

% bullet clusters 
% + defelction/lensing 

\subsection*{Acknowledgments}  
We thank Prof. M. Milgrom for his stimulating criticism at the earlier stages of this work, 
Prof.  A. Yahalom and M. Wright for their genuine interest, Dr. H. Zhao for his comments on our previous paper, 
and Dr. I. Banik for his question about the compatibility of our corrections with the Cassini data.  
We also thank Ilya Kholodnyi for his help with editing the English of the paper.

%\bibliographystyle{spmpsci}

%\bibliographystyle{ws-procs961x669}
%\bibliography{ws-pro-sample}

\begin{thebibliography}{00}

\bibitem{ameli1} G. Amelino-Camelia, Introduction to Quantum-Gravity phenomenology, 
in: {\em Planck Scale Effects in Astrophysics and Cosmology}, 
eds. J. Kowalski-Glikman, G. Amelino-Camelia, 
 Lecture Notes in Physics, vol 669. Springer, Berlin, Heidelberg 2005,  %https://doi.org/10.1007/1137730
59; %-100; 	
{\tt arXiv:gr-qc/0412136}. 

\bibitem{menezes} G. Menezes, 
Quantum gravity phenomenology from the perspective of quantum general relativity and quadratic gravity, 
Class. Quantum Grav. {\bf 40} (2023) 235007; {\tt arXiv:2305.19517}. %[gr-qc]

\bibitem{addazi} A. Addazi, e.a. 
Quantum gravity phenomenology at the dawn of the multi-messenger era - a  review, 
Prog. Part.  Nucl. Phys. {\bf 125}  (2022) 103948. 
% 10.1016/j.ppnp.2022.103948

\bibitem{alonso} N. Huggett, N. Linnemann and  M. Schneider, 
Quantum Gravity in a Laboratory? {\tt 	arXiv:2205.09013}. 

%A. Alonso-Serrano and M. Li\v{s}ka, 
%Quantum phenomenological gravitational dynamics: a general view from thermodynamics of spacetime, 
%JHEP, 2020 (2020)  196. 
%Quantum gravity phenomenology from thermodynamics of spacetime, 
%The Sixteenth Marcel Grossmann Meeting,  462-478 (2023). 

\bibitem{freidel} L. Freidel, J. Kowalski-Glikman, R.G. Leigh, and D. Minic, 
Quantum gravity phenomenology in the infrared, 
Int. J. Mod. Phys. {\bf D30} (2021) 2141002.  

%metric formulation  \cite{ikm1,ikm2,ikm3,ikm4} 
\bibitem{ikm1} I. V. Kanatchikov, 
 From the DeDonder-Weyl Hamiltonian formalism to quantization of gravity, 
in: {\em Current topics in mathematical cosmology}, 
eds. M. Rainer and H.-J. Schmidt, World Scientific 1998, 472; %-482; 
{\tt 	arXiv:gr-qc/9810076}.
 \bibitem{ikm2} I. V. Kanatchikov, 
  Quantization of gravity: yet another way, 
in: {\em Coherent states, quantization and gravity}, eds. M. Schlichenmaier, 
 A. Strasburger, S. T. Ali and A. Odzijewicz, 
 Warsaw University Press, Warsaw 2001, 189; % - 197; % pp. 189-197.
{\tt arXiv:gr-qc/9912094}. 
 \bibitem{ikm3} I. V. Kanatchikov, 
 Precanonical perspective in quantum gravity, 
Nucl. Phys. Proc. Suppl. %Suppl. 
{\bf 88} (2000) 326; %-330
 %doi:10.1016/S0920-5632(00)00795-7, 
\!{\tt arXiv:gr-qc/0004066}.  
  \bibitem{ikm4} I. V. Kanatchikov, 
 Precanonical quantum gravity: quantization without the space-time decomposition, 
Int. J. Theor. Phys.  {\bf 40} (2001) 1121;    %-1149
{\tt arXiv:gr-qc/0012074}. 
%doi 10.1023/A:1017557603606
  
  
  %Palatini vielbein formulation \cite{ikv1,ikv2,ikv3,ikv4,ikv5}. 
\bibitem{ikv1} 
  %\bibitem{pqg-vielbein}   
I. V. Kanatchikov, 
 { On precanonical quantization of \ gravity in spin connection variables}, 
{\rm AIP Conf. Proc. } {\bf 1514} (2012) 73, % (2012), % 73-76,  
{\tt arXiv:1212.6963}.
\bibitem{ikv2} 
I. V. Kanatchikov, 
{  De Donder-Weyl Hamiltonian formulation and precanonical quantization of \ vielbein gravity}, 
{\rm J. Phys. Conf. Ser.} {\bf 442} (2013) 012041, 
{\tt arXiv:1302.2610}. 
\bibitem{ikv3} 
I. V. Kanatchikov, 
 {  On precanonical quantization of \ gravity}, 
 %%{\rm Nonlin. Phenom. Complex Sys. (NPCS)} 
 NPCS {\bf 17} (2014) 372, %(2014). %372-376, 
 {\tt arXiv:1407.3101}. 
\bibitem{ikv4} I. V. Kanatchikov, 
 Ehrenfest theorem in precanonical quantization of fields and gravity,
in {\em Proc. of the Fourteenth Marcel Grossmann Meeting on General Relativity}, 
eds.  M. Bianchi, R. T. Jantzen, R. Ruffini, World Scientific (2018)  2828; %-2835; 
 %https://doi.org/10.1142/9789813226609_0352
{\tt arXiv:1602.01083}. 
\bibitem{ikv5} I.V. Kanatchikov, 
 On the ``spin connection foam" picture of quantum gravity from precanonical quantization, 
in {\em Proc. of the Fourteenth Marcel Grossmann Meeting on General Relativity}, 
eds.  M. Bianchi, R. T. Jantzen, R. Ruffini, World Scientific (2018)  3907; %-3915; 
 %https://doi.org/10.1142/9789813226609_0519
{\tt arXiv:1512.09137}.   



%\bibitem{rund} H. Rund, {\em Hamilton-Jacobi theory in the calculus of variations: its role in mathematics and physics}, 
%ISBN 10: 0442070993  ISBN 13: 9780442070991
%D. Van Nostrand Co.  (1966). 


%%10 new
\bibitem{kk} I. V. Kanatchikov and V. A. Kholodnyi, 
The Milgromian acceleration and the cosmological constant from precanonical quantum gravity, 
 in: {\em Geometric Methods in Physics XL}, 
 eds. P. Kielanowski, D. Beltita, A. Dobrogowska, T. Goli\'nski, 
 Springer (2024) 393; %–401;  
{\tt arXiv:2311.05525}.   
%doi: 10.1007/978-3-031-62407-0_26



%% DW 11, 12 
\bibitem{dedonder} Th. De Donder, Th\'eorie invariantive du calcul des variations, Gauthier‐Villars, Paris (1930). %% or 1935 nouv. éd.

\bibitem{kastrup} H. Kastrup, Canonical theories of Lagrangian dynamical systems in physics, 
Phys. Rep. {\bf 101} (1983) 1. %-167. %Issues 1–2, December 1983, Pages 1-167

%%13-16 

\bibitem{mybr1} I. V. Kanatchikov, 
On the canonical structure of De Donder-Weyl covariant Hamiltonian 
formulation of field theory I, %Graded Poisson brackets and equations of motion, 
{\tt arXiv:hep-th/9312162}. 
 
\bibitem{mybr2} I. Kanatchikov,   
 Canonical structure of classical field theory in the polymomentum phase space, 
 Rep. Math. Phys.  {\bf 41} (1998) 49;   
{\tt arXiv:hep-th/9709229}. 
 
\bibitem{mybr3} 
I. Kanatchikov, 	
  On field theoretic generalizations of a Poisson algebra,  
 Rep. Math. Phys. {\bf 40} (1997) 225;   
 {\tt  arXiv:hep-th/9710069}.

\bibitem{ik5} I. V. Kanatchikov, 
 Geometric (pre)quantization in the polysymplectic approach to field theory, 
in: {\em Differential geometry and its applications,}   
 %Proc. Conf., Opava (Czech Republic), August 27–31, 2001
 eds. O. Kowalski, D. Krupka and J. Slovak, 
 Silesian University, Opava (2001), 
 309; %–321; 
{\tt arXiv:hep-th/0112263}. 

\bibitem{mydirac} 
I.V. Kanatchikov,
 {  On a generalization of the Dirac bracket in the De Donder-Weyl Hamiltonian formalism, } 
in: {\em Differential Geometry and its Applications, } 
 %Proc. 10th Int. Conf. on Diff. Geom. \& Appl., Olomouc, August 2007, 
 ed. Kowalski O., Krupka D., Krupkov\'a O. and Slov\'ak J., 
 World Scientific, Singapore 2008, 615; %-625, 
{\tt arXiv:0807.3127}. 


%precanonical quantization \cite{ik5e,ik2,ik3,ik4,ik5}

\bibitem{ik5e} I. V. Kanatchikov, 
 Ehrenfest theorem in precanonical quantization, 
J. Geom. Symmetry Phys. {\bf 37} (2015) 43; %-66
 %doi:10.7546/jgsp-37-2015-43-66, 
{\tt arXiv:1501.00480}.  
\bibitem{ik2} I. Kanatchikov, 
 Towards the Born-Weyl quantization of fields, 
Int. J. Theor. Phys. {\bf 37} (1998) 333; %%!-342
{\tt arXiv:quant-ph/9712058}.  
\bibitem{ik3} I. V. Kanatchikov, 
 DeDonder-Weyl theory and a hypercomplex extension of quantum mechanics to field theory, 
Rept. Math. Phys. {\bf 43} (1999) 157;  %-170
 %doi:10.1016/S0034-4877(99)80024-X,
{\tt arXiv:hep-th/9810165}.  
\bibitem{ik4} I. V. Kanatchikov, 
 On Quantization of field theories in polymomentum variables, 
AIP Conf. Proc. {\bf 453} (1998) 356; %-367,1998 % doi:10.1063/1.57105, 
{\tt arXiv:hep-th/9811016}.






%\cite{iky3,iks1,iks2,iksc1,iksc2,iksc3}

%Scalar 
\bibitem{iks1} I. V. Kanatchikov, 
Precanonical quantization and the Schrödinger wave functional revisited,
Adv. Theor. Math. Phys. {\bf 18} (2014) 1249; %-1265
{\tt arXiv:1112.5801}.
\bibitem{iks2} I. V. Kanatchikov, 
On the precanonical structure of the Schr\"odinger wave functional,
Adv. Theor. Math. Phys. {\bf 20} (2016) 1377; %-1396
{\tt arXiv:1312.4518}. 

%Scalar curved 
\bibitem{iksc1}
I. V. Kanatchikov, 
 Schr\"odinger functional of a quantum scalar field in static space-times from precanonical quantization, 
Int. J. Geom. Meth. Mod. Phys. {\bf 16} (2019) 1950017, 
{\tt arXiv:1810.09968}.   
\bibitem{iksc2}
I. V. Kanatchikov, 
 Precanonical structure of the Schr\"odinger wave functional in curved space-time,
Symmetry {\bf 11} (2019) 1413;  %https://doi.org/10.3390/sym11111413
{\tt  arXiv:1812.11264}.
\bibitem{iksc3} 
I. V. Kanatchikov, 
 On the precanonical structure of the Schr\"odinger wave functional in curved space-time, 
Acta Phys. Polon. B Proc. Suppl. {\bf 13} (2020) 313;  
{\tt arXiv:1912.07401}.  

%YM
\bibitem{iky1} I. V. Kanatchikov, 
Precanonical quantization of Yang-Mills fields and the functional Schroedinger representation, 
Rep. Math. Phys. {\bf 53} (2004) 181;  %-193
%doi:10.1016/S0034-4877(04)90011-0, 
{\tt arXiv:hep-th/0301001}. 
\bibitem{iky3} I. V. Kanatchikov, 
Schr\"odinger wave functional in quantum Yang-Mills theory from precanonical quantization, 
Rep. Math. Phys. {\bf 82} (2018) 373;  
%doi:10.1016/S0034-4877(19)30008-4, 
{\tt arXiv:1805.05279}. 

\bibitem{my-ymmg} I. V. Kanatchikov, 
 On the spectrum of DW Hamiltonian of quantum SU(2) gauge field,
Int. J. Geom. Meth. Mod. Phys. {\bf 14} (2017) 1750123, 
%doi:10.1142/S0219887817501237,
{\tt arXiv:1706.01766}. 


%\bibitem{meinrenken} E. Meinrenken, {\em Clifford algebras and Lie theory}, Springer-Verlag (2013). 
%Berlin Heidelberg (2013).

%\bibitem{volterra} A. Slav\'\i k, {\em Product integration, its history and applications}, Matfyzpress, Prague (2007).  
 %{http://www.karlin.mff.cuni.cz/~slavik/product/product_integration.pdf}
%V. Volterra and B. Hostinsk\'y, {\em Op\'erations infinit\'esimales lin\'eaires. 
%Applications aux \'equations diff\'erentielles et fonctionnelles.} Paris, Gauthier-Villars (1938).



 %\bibitem{kk23} I. V. Kanatchikov and V. A. Kholodnyi, work in progress. 
 
 %35
 \bibitem{my-mink} I. V. Kanatchikov, 
 The quantum waves of Minkowski spacetime and the minimal acceleration from precanonical quantum gravity, 
J. Phys.: Conf. Ser. {\bf 2533} (2023) 012037;  
{\tt arXiv:2308.08738}.


\bibitem{iktp1} I. V. Kanatchikov, 
Towards precanonical quantum teleparallel gravity,  
{\tt arXiv:2302.10695}.

%37
\bibitem{iktp2} I. V. Kanatchikov, 
The De Donder-Weyl Hamiltonian formulation of TEGR and its quantization, 
{\tt arXiv:2308.10052}.  
 
 
 
 
 %38 
\bibitem{mond83} M. Milgrom, 
A modification of the Newtonian dynamics as a possible alternative to the hidden mass hypothesis, 
Astrophysical Journal, {\bf 270} (1983) 365. %-370.

 %39
\bibitem{mond} M. Milgrom, MOND theory, Can. J. Phys. {\bf 93} (2015) 107; {\tt arXiv:1404.7661}. 

 %40 
\bibitem{mond6} M. Milgrom,  
The $a_0$ $-$ cosmology connection in MOND, 
{\tt arXiv:2001.09729}. 

 %41
 \bibitem{mond-th}
M. Milgrom, MOND—theoretical aspects, 
New Astronomy Reviews, {\bf 46} (2022) 741. %-753.
%Volume 46, Issue 12, November 2002, Pages 741-753

%M. Milgrom,
%Generalizations of Quasilinear MOND (QUMOND) 
%Phys. Rev. D 108, 084005 (2023), 
%	arXiv:2305.01589 .
	

%42
\bibitem{mond5} 
B. Famaey and S. McGaugh, 
 Modified Newtonian Dynamics (MOND): Observational phenomenology and relativistic extensions, 
Living Rev. Rel. {\bf 15} (2012) 10; 
{\tt arXiv:1112.3960}. % [astro-ph.CO]ppwm  

%43
\bibitem{cornell1}
E. Eichten, K. Gottfried, T. Kinoshita, J. Kogut, K.D. Lane, and T.M. Yan, 
Spectrum of charmed quark-antiquark bound states,
Phys. Rev. Lett. {\bf 34} (1975) 369, [Erratum: Phys. Rev. Lett. {\bf 36} (1976) 1276.
% E. Eichten, K. Gottfried, T. Kinoshita, K. D. Lane and T.-M. Yan, Phys. Rev. D
%17, 3090 (1978), [Erratum: Phys.Rev.D 21, 313 (1980)].
% E. J. Eichten, K. Lane and C. Quigg, Phys. Rev. Lett. 89, 162002 (2002).

\bibitem{grum} D. Grumiller, Model for gravity at large distances, 
Phys. Rev. Lett. {\bf 105} (2010) 211303.  

\bibitem{grum2} D. Grumiller and F. Preis,
Rindler force at large distances,  Int. J. Mod. Phys. {\bf D20} (2011) 2761; -2766
	{\tt arXiv:1107.2373}. 

\bibitem{mannheim} P.D. Mannheim and D. Kazanas, Exact vacuum solution to conformal Weyl gravity and 
galactic rotation curves,  ApJ. {\bf 342} (1989)  635. %–638.


\bibitem{conf-modesto} L. Modesto, T. Zhou and Q. Li,
Geometric origin of the galaxies’ dark side, 
Universe {\bf 10} (2024) 19. 
 % https://doi.org/10.3390/universe10010019


%44 
\bibitem{milg99}
M. Milgrom, The modified dynamics as a vacuum effect, Phys. Lett. {\bf A 253} (1999) 273, 
{\tt
 arXiv:astro-ph/9805346}.

%45 
 \bibitem{pikh10}
 P. V. Pikhitsa, MOND reveals the thermodynamics of gravity, {\tt arXiv:1010.0318}.

%46
\bibitem{klink12} 
 F.R. Klinkhamer,  Entropic-gravity derivation of MOND, 
 Mod. Phys. Lett. {\bf A27} (2012) 1250056; {\tt arXiv:1201.4160}. 


%M. Cadoni and M. Tuveri, 	
%	Galactic dynamics and long-range quantum gravity,

%\bibitem{mie1903} G. Mie, Zur kinetischen Theorie der einatomigen K\"orper, 
% Ann. der Physik {\bf 316} (1903) 657. %–697. https://doi.org/10.1002/andp.19033160802 (1903).


%47 
\bibitem{deur16}
A. Deur, Stanley J. Brodsky, G. F. de Teramond,  
The QCD running coupling, 
 Prog. Part. Nucl. Phys. {\bf 90} (2016) 1; 
  %-74; 
{\tt  arXiv:1604.08082}. % [hep-ph]

%48
\bibitem{deur23}
A. Deur, S.J. Brodsky, C.D. Roberts, 
QCD running couplings and effective charges, 
Prog. Part. Nucl. Phys. {\bf 134} (2024) 104081; 
{\tt 	arXiv:2303.00723}. 

%49
\bibitem{sanctis24} 
M. De Sanctis, 
Phenomenological exploration of the strong coupling
constant in the perturbative and nonperturbative regions,   
  %%alphaS (Q=0) approx 2 + refs
{\tt arXiv:2410.21628}. 
 
% 50 
\bibitem{kk25} V.A. Kholodnyi and I.V. Kanatchikov, work in progress. 

\bibitem{subatto1} V. Bl\-{u}ms,  M. Piotrowski, M. I. Hussain, B. G. Norton, S. C. Connell, S. Gensemer, M. Lobino, E. W. Streed,  
A single-atom 3D sub-attonewton force sensor, 
Sci. Adv. {\bf 4} (2018) aao4453; %DOI: 10.1126/sciadv.aao4453
	{\tt arXiv:1703.06561}.

\bibitem{subatto2} Z. Liu, Y. Wei, L. Chen, Ji Li, S. Dai, F. Zhou, M. Feng, 
Phonon-laser ultrasensitive force sensor, Phys. Rev. Appl. {\bf 16} (2021) 044007;  
	{\tt arXiv:2110.01146}. 
	
	\bibitem{atto}  
T. M. Fuchs, D. G. Uitenbroek, J. Plugge, N. van Halteren, J.-P. van Soest, A. Vinante, H. Ulbricht, and T. H. Oosterkamp, 
Measuring gravity with milligram levitated masses, Sci. Adv. {\bf 10} (2024) eadk2949; 
	{\tt arXiv:2303.03545}. 
%%https://physicsworld.com/a/getting-closer-to-measuring-quantum-gravity/

%51
%M31: 
\bibitem{m31-24}
X. Zhang, B. Chen, P. Chen, J. Sun, and Z. Tian, 
The rotation curve and mass distribution of M31, MNRAS {\bf 528} (2024) 2653; 
%–2666, https://doi.org/10.1093/mnras/stae025
{\tt 	arXiv:2401.01517}. % [astro-ph.GA]

%52
\bibitem{m31-06}
C. Carignan, L. Chemin, W. Huchtmeier, and F.J. Lockman, 	
Extended $HI$ rotation curve and mass distribution of M31, %up to 35kpc, 20% error. fig2
%March 2006The Astrophysical Journal 641(2)
ApJ. {\bf 641} (2006)  L109; %L112; 
{\tt arXiv:astro-ph/0603143}. 

%MW: 	
\bibitem{mw-14}
P. Bhattacharjee, S. Chaudhury, and S. Kundu,
	Rotation curve of the Milky Way out to 200 kpc, 
%The Astrophysical Journal, Volume 785, Number 1
%Citation Pijushpani Bhattacharjee et al 2014 
ApJ. {\bf 785} (2014) 3; 
%DOI 10.1088/0004-637X/785/1/63
	{\tt arXiv:1310.2659}. 
 
\bibitem{clusters}
A.O. Hodson and H. Zhao, Generalizing MOND to explain the missing mass in galaxy clusters, 
%Astronomy $\&$ Astrophysics %
A{\&}A {\bf 598} (2017) A127; %	https://doi.org/10.1051/0004-6361/201629358
{\tt  arXiv:1701.03369}. % [astro-ph.CO]
 
 \bibitem{mcgaugh24}  
 S.S. McGaugh, J.M. Schombert, F. Lelli and J. Franck, 
 Accelerated structure formation: The early emergence of massive galaxies and clusters of galaxies, 
 ApJ {\bf 976} (2024) 13;  
 	{\tt arXiv:2406.17930}. 
%The Astrophysical Journal, Volume 976, Number 1
%Citation Stacy S. McGaugh et al 2024 ApJ 976 13
%DOI 10.3847/1538-4357/ad834d
 
%up to 1Mpc
\bibitem{1mpc}
T. Mistele, S. McGaugh, F. Lelli, J. Schombert, and P. Li, 
 Indefinitely flat circular velocities and the baryonic Tully-Fisher relation from weak lensing,  
ApJ. Lett.  {\bf 969} (2024) L3;
{\tt arXiv:2406.09685}. % [astro-ph.GA]. 
	
	\bigskip 
	%\pagebreak 
	 \small
{\!\!\!\!\!\!\!\!}After this Proceedings paper was submitted in December 2024, the results were further 

{\!\!\!\!\!\!\!\!}developed and published in: 
\medskip 

\bibitem{epl25} I.V. Kanatchikov and V.A. Kholodnyi, 
Observable signatures of precanonical quantum gravity,
  {\it EPL} {\bf 150} 59002 (2025).
  %, \url{http://dx.doi.org/10.1209/0295-5075/addcdb}.

\bibitem{dice24} I.V. Kanatchikov and V.A. Kholodnyi,
Dark matter and dark energy as manifestations of quantum spin connection foam, 
%Journal of Physics: Conference Series, Volume 3017, Eleventh International Workshop on Decoherence, Information, Complexity and Entropy (DICE 2024) 15/09/2024 - 20/09/2024 Castiglioncello, Italy
 {\it J. Phys.: Conf. Ser.} {\bf 3017} 012031 (2025).
 %, \url{http://dx.doi.org/10.1088/1742-6596/3017/1/012031}. 

\bibitem{mpla25} I.V. Kanatchikov, V.A. Kholodnyi, J. Kozicki and M.E. Pietrzyk,
 Modified Newtonian Dynamics from higher moments of quantum spin connection in precanonically quantized gravity,
Mod. Phys. Lett. A {\bf 41} 2541012 (2026). 




\end{thebibliography}

% \vfill
%\pagebreak

\end{document}